# Magnetoelectric Control of Helical Light Emission in a Moiré Chern Magnet

Eric Anderson[1*], Heonjoon Park[1], Kaijie Yang[2], Jiaqi Cai[1], Takashi Taniguchi[3], Kenji Watanabe[4], Liang Fu[5], Ting Cao[2], Di Xiao[2,1†], and Xiaodong Xu[1,2‡]

[1]Department of Physics, University of Washington, Seattle, Washington 98195, USA
[2]Department of Materials Science and Engineering, University of Washington, Seattle, Washington 98195, USA
[3]Research Center for Materials Nanoarchitectonics, National Institute for Materials Science, 1-1 Namiki, Tsukuba 305-0044, Japan
[4]Research Center for Electronic and Optical Materials, National Institute for Materials Science, 1-1 Namiki, Tsukuba 305-0044, Japan
[5]Department of Physics, Massachusetts Institute of Technology, Cambridge, Massachusetts 02139, USA

[*]Present address: Université Paris-Saclay, CNRS, Centre de Nanosciences et de Nanotechnologies (C2N), Palaiseau 91120, France.
[†]Contact author: dixiao@uw.edu
[‡]Contact author: xuxd@uw.edu

**ABSTRACT**. Magnetoelectric effects and their coupling to light helicity are important for both fundamental science and applications in sensing, communication, and data storage. Traditional approaches require complex device architectures, involving separate spin-injection, ferromagnetic, and optically active layers. Recently, the emergence of 2D semiconductor moiré superlattices with flat Chern bands and strong light-matter interactions has established a simple yet powerful platform for exploring the coupling between photon, electron, and spin degrees of freedom. Here, we report efficient current control of spontaneous ferromagnetism and associated helicity of light emission in moiré MoTe$_2$ bilayer – a system which hosts a rich variety of topological phases, including newly discovered zero-field fractional Chern insulators. We show that the current control is effective over a wide range of doping of the first moiré Chern band, implying the uniformity of the Berry curvature distribution over the flat band. By setting the system into the anomalous Hall metal phase, a current as small as 10nA is sufficient to switch the magnetic order, a substantial improvement over both conventional spin torque architectures and other moiré systems. The realized current control of ferromagnetism leads to continuous tuning of trion photoluminescence helicity from left to right circular via spin/valley Hall torque at zero magnetic field. Our results pave the way for topological opto-spintronics based on semiconductors with synthetic flat Chern bands.

## I. INTRODUCTION

Moiré superlattice engineering has been proven to be an effective approach for creating new phases of matter with exotic physical properties. [1–3] A striking example is the recent discovery of the fractional quantum anomalous Hall (FQAH) effect in twisted MoTe$_2$ bilayer [4–10] (tMoTe$_2$) and rhombohedral-stacked multilayer graphene, [11] realizing long-sought zero-field fractional Chern insulators. [12–18] Besides hosting the FQAH effect, the semiconducting transition metal dichalcogenide (TMD) tMoTe$_2$ is a direct bandgap semiconductor, with spin-valley locking and a strong excitonic response. [19–21] These properties have enabled optical probing of the Chern insulator [4] and putative composite Fermi liquid [10] states via measurement of the intensity and helicity of trion photoluminescence. In addition, both electric field and doping have been demonstrated to be effective controls of the spontaneous ferromagnetic order in tMoTe$_2$, [21] enabling powerful magnetoelectric manipulation.

The physical properties of tMoTe$_2$ mentioned above offer a new means to manipulate intrinsic magnetism – and thus light emission helicity, which is tightly linked to spin/valley polarization – via electric current. Current control of light emission helicity is highly sought after, as it involves coupling the relevant degrees of freedom in electronics, spintronics, and photonics. This functionality has recently been achieved in complex multilayer architectures. [22] For these devices, charge current in a layer with strong spin-orbit coupling generates a spin current, and the resulting spin-orbit torque enables magnetization switching in an adjacent ferromagnetic layer. This further allows for injection of spin-polarized carriers into a separate optically active layer. However, the degree of circular polarization of emitted light is limited, largely due to the spatial separation between the spin current generating and the optically active layers.

Here, we realized spin/valley Hall torque driven magnetization switching and current controlled helical light emission in tMoTe$_2$. Remarkably, the degree of circular polarization can be continuously tuned, and reaches nearly perfect circular polarization at a small applied current. This is achieved via three unusual properties of the system. First, the necessary ingredients – spin torque control of magnetic order and spin polarization-dependent helical emission - occur within a single direct bandgap semiconductor system, overcoming the challenge of multilayer structures. Second, flat Chern bands with nearly uniform Berry curvature and spin-valley locking effects enable efficient spin/valley polarization control over a large doping range. Third, tMoTe$_2$ hosts rich and electrically tunable quantum phases. Tuning the system into an anomalous Hall metal phase [23] – i.e., a metallic phase with anomalous Hall effect – increases the efficiency of current control. Below, we present results from two tMoTe$_2$ devices, D(3.7°) and D(3.9°), with twist angles of 3.7° and 3.9°. Both devices host integer and fractional QAH states, as detailed in prior reports. [6,8]

## II. RESULTS AND DISCUSSION

### A. Anomalous Hall metal

We begin with an exploration of the anomalous Hall metal phase (AHM). Figure 1(a) shows a schematic of the dual gated device and an atomic force microscopy (AFM) image of D(3.9°). The clean surface implies a high device quality. Figure 1(b) is a plot of longitudinal resistance ($R_{xx}$) as a function of doping and out-of-plane electric field (or displacement field) $D/\varepsilon_0$. Data are taken at a temperature of 100 mK in D(3.7°) and symmetrized at magnetic field $\mu_0H=\pm100$ mT. As identified in a prior report, [6] there are QAH and FQAH phases near hole fillings $\nu=-1$ and $\nu=-2/3$, respectively, with suppressed $R_{xx}$. Here, $\nu$ is the filling factor, denoting the number of carriers per moiré unit cell. Between $\nu=-1$ and -2/3, $R_{xx}$ is also suppressed, while $R_{xy}$ is finite but smaller than $R_{xx}$.

A measurement of reflective magnetic circular dichroism (RMCD) at 25 mT in the same region of phase space is shown in Figure 1(c). All optical measurements are taken at 1.6 K unless otherwise indicated. The small applied magnetic field is to suppress spin fluctuations due to the doping sweep (see FIG. S1). As RMCD is sensitive to spin/valley polarization, the measurement reveals the ferromagnetic phase space versus $v$ and $D/\varepsilon_0$. Figure 1(d) illustrates $R_{xx}$ and $R_{xy}$ versus $\mu_0H$ at hole filling $v = -0.78$ and $D/\varepsilon_0 = 0$. As shown by the RMCD measurement in Fig. 1(c), the system is in a spin-valley polarized (ferromagnetic) regime under these conditions. $R_{xy}$ shows clear hysteresis and is on the order of a few k$\Omega$, while $R_{xx}$ is nearly constant at ~10 k$\Omega$, with a slight increase near the critical field $H_C \approx \pm20$ mT where the sign of $R_{xy}$ flips. Thus, the anomalous Hall effect is observed at this partial filling of the first moiré Chern band.

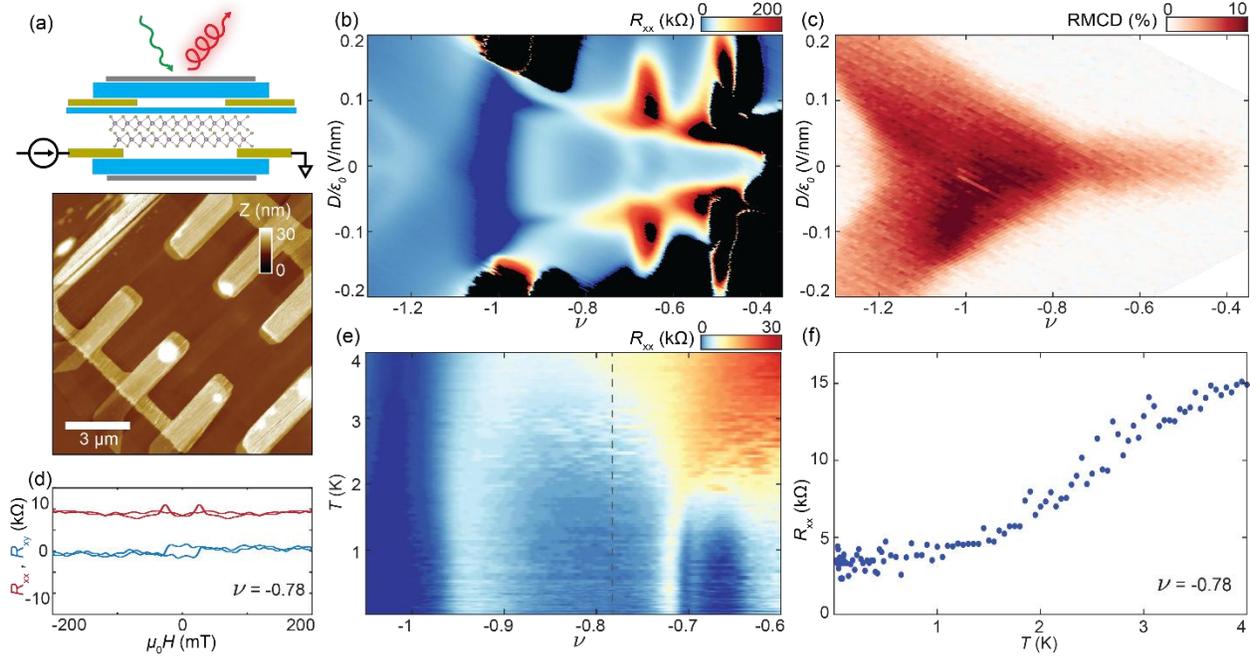

FIG. 1. Anomalous Hall metal in moiré tMoTe$_2$. (a) Top: schematic of device cross section, with graphite gates in grey, BN dielectric layers in blue, and Pt contacts in yellow. Top Pt gates are used to improve contact transparency, and current is injected via the sample contacts. Optical probes can be used to reveal spontaneous spin/valley polarization with current controls. Bottom: atomic force microscopy (AFM) image of device D(3.9°), with contacts for current injection visible. (b) Longitudinal resistance $R_{xx}$ vs moiré filling $\nu$ and displacement field $D/\varepsilon_0$ in D(3.7°), consistent with previous measurements. Resistance in black regions is too high to be reliably measured. $R_{xx}$ is symmetrized at $|\mu_0H| = 200$ mT. (c) As in (b), but measuring reflective magnetic circular dichroism (RMCD). Region with finite RMCD signal is ferromagnetic. Sample was measured at $\mu_0H = 25$ mT to prevent domain flipping as gates were swept. (d) Longitudinal ($R_{xx}$) and Hall ($R_{xy}$) resistance vs magnetic field in the AHM metal regime. $R_{xy}$ hysteresis and nonvanishing signal at 0T are signatures of the anomalous Hall effect. (e) Temperature dependence of $R_{xx}$ vs $\nu$ in D(3.9°). (f) Temperature dependence of $R_{xx}$ at $\nu = -0.78$, extracted along the dotted line in (e). Increasing resistance with temperature indicates metallic behavior.

We further performed temperature dependent measurements. Figure 1(e) shows a measurement of $R_{xx}$ versus filling and temperature in D(3.9°). Around $v = -1$ and $v = -2/3$, $R_{xx}$ remains nearly vanishing at low temperature and rises as temperature is increased, when the thermal activation of the carriers over the Chern gap dominates. Between -1 and -2/3, $R_{xx}$ continuously increases with temperature. Taking a linecut at $v = -0.78$ (dashed line), $R_{xx}$ increases from about 4 k$\Omega$ to 15 k$\Omega$ as temperature increases from 100 mK to 4 K (Fig. 1(f)), suggesting a metallic phase. In addition, previous measurements of trion photoluminescence versus doping [4,10] do not show suppression of the luminescence intensity in this region of phase space, in contrast to the suppression observed at Chern insulator states. This is consistent with the compressible nature of a metallic phase. Taken together with $R_{xx}$ being several times larger than $R_{xy}$, the data demonstrate the system is an AHM over this region of doping – i.e. a spontaneous ferromagnetic metal phase exhibiting anomalous Hall signal in a partially filled Chern band.

## B. Current control of spin/valley polarization

Next, we examine current control of ferromagnetism, focusing on the anomalous Hall metal phase. We employ RMCD to probe the ferromagnetic order. For instance, Fig. 2(a) shows the RMCD signal as $\mu_0H$ is cycled at $v = -0.77$ and $D/\varepsilon_0 = 0$, i.e. in the AHM. The clear RMCD hysteresis with finite signal at zero magnetic field is consistent with the observed anomalous Hall signal in Fig. 1(d). To determine the effect of injected charge current ($I$) on the spin/valley polarization, we first took a spatial map of the RMCD signal over the entire sample at $I = 0$. Figure 2(b) shows this map for the same $v$ and $D/\varepsilon_0$ as in Fig. 2(a), measured at $\mu_0H = 0$ with magnetization initialized to point up. Finite, positive RMCD signal exists across the entire sample area, with contacts visible as regions of decreased RMCD (outlined with grey dashed lines). This is the expected behavior for a homogeneous sample tuned to the FM phase.

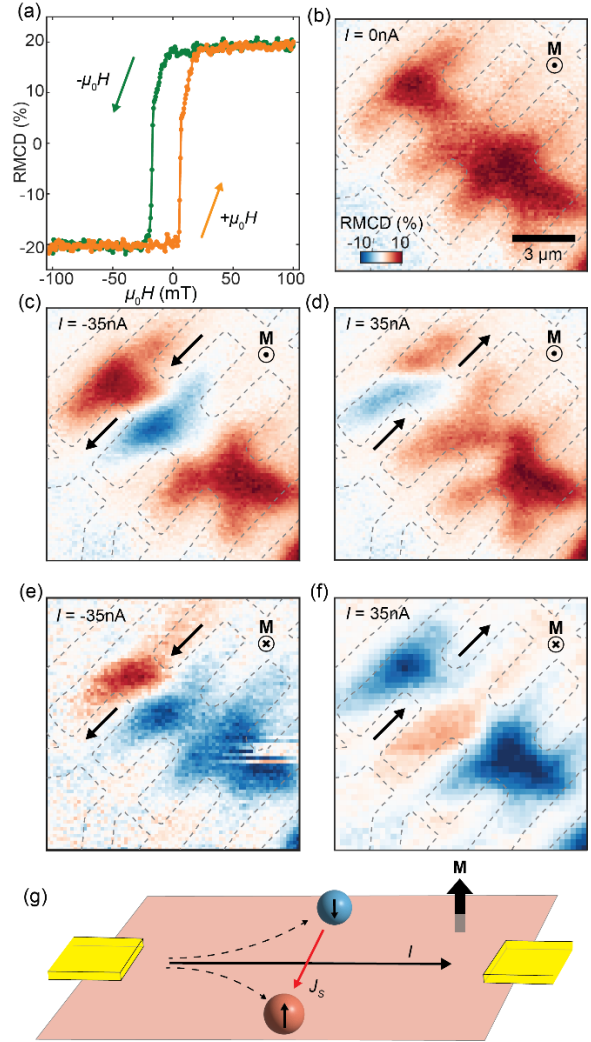

FIG. 2. Transverse spin and charge current in a ferromagnetic metal. (a) RMCD signal as a function of $\mu_0H$ swept down and up. RMCD hysteresis demonstrates ferromagnetic order. (b) RMCD spatial map in the anomalous Hall metal regime ($v = -0.77$) with magnetization pointing up. Grey dashed lines indicate Pt contact area. (c) RMCD map under a current of $I = -35$ nA (indicated by the arrows) flowing between the upper-left contacts of the device. The sign change in RMCD implies the reversal of spin/valley polarization. (d) RMCD map with the current direction reversed. The sign-reversed domain appears on the opposite side of the current channel. (e) RMCD map with magnetization pointing down and $I = -35$ nA, as in (c). Despite the reversed magnetization, domains signs on either side of the current channel remain consistent with (c). (f) RMCD map with the reversed current direction and magnetization pointing down. Signs of domains on either side of the current channel match those in (d). (g) Schematic of the spin/valley Hall effect.

In stark contrast, when an RMCD spatial map is taken under the same conditions as Fig. 2(b), but with current flowing between the two contacts on the top left, clear domains with negative RMCD signal are observed (Figs. 2(c)&(d)). The current injection configuration is discussed further in Methods. The opposite sign of the RMCD signal in these domains compared to the rest of the sample area demonstrates a current-driven spin/valley polarization reversal. The spatial position of the domain with reversed spin/valley polarization can be controlled by the current flow direction. For $I$ = -35nA (Fig. 2(c)), this negative domain with magnetization pointing down (blue) is to the lower right of the current channel. In contrast, for current flowing in the opposite direction (Fig. 2(d)), the negative domain is on the upper left side.

We found that the spatial pattern of current-induced polarization reversed domains is independent of the initial magnetization orientation. The RMCD map in Fig. 2(e) is taken with the magnetization initialized to point down, as evidenced by the opposite sign of the RMCD signal on the bottom right of the map, far from the current channel. Comparing this map to Fig. 2(c) taken at $I$ = -35 nA, although the initial magnetization direction is reversed, the signs of the domains on either side of the current channel are the same – i.e., positive on the upper left and negative on the lower right. Similar behavior is observed when comparing Figs. 2(d)&(f) - for the same $I$ = 35 nA current, the signs of the domains on either side of the current channel match, even though the initial magnetization directions are reversed. Varying the source and drain pin configuration changes the current channel and thus the locations of the domains (FIG. S2), but the same qualitative behavior remains. Similar RMCD maps taken at selected current values (FIG. S3) show that current-induced polarization reversed domains are present for values of $I$ as low as 10 nA.

The above measurements demonstrate that the magnetic domain pattern is only determined by the current flow direction. This results from the mechanism for current-induced magnetization flipping – spin/valley Hall torque, as shown in Fig. 2(g). The spin/valley Hall torque switching of magnetic order has been observed in MoTe$_2$/WSe$_2$ heterobilayer superlattices, but only near full filling of the moiré valence band (or $v$ = -1). [24,25] The observation was attributed to the large Berry curvature near $v$ = -1, [3,26,27] where the QAH effect was observed. [28] Our results demonstrate the same mechanism works in tMoTe$_2$.

A major distinction in tMoTe$_2$, however, lies in two unique aspects of the system. The first is the low current density required for magnetization switching in the anomalous Hall metal phase. At approximately ~$10^2$ A/cm$^2$, the switching current is nearly an order of magnitude lower compared to other moiré systems. [24,25] This is most likely due to the fact that in other systems the FM phase only exists close to the insulating state, which has a relatively large spin/valley energy splitting. In contrast, in the AHM phase, the spin splitting is decreased due to carrier screening and/or partial valley polarization. The second aspect is that tMoTe$_2$ exhibits this current control behavior over a much larger region of phase space. Although in the above discussion we focus on the AHM, efficient current switching is observed over a broad range of fillings of the first moiré Chern band. As shown in FIG. S4, the sign of RMCD changes with applied current not only near $v$ = -0.77, but over the entire FM phase space for hole fillings below $v$ = -1, as well as in the wings of FM phase above $v$ = -1 with large electric field. Because the transverse spin current arises from the Berry curvature of the moiré bands, this broad range demonstrates that the Berry curvature remains substantial far from the $v$ = -1 band edge, which is consistent with the nearly uniform Berry curvature condition needed for the observed FCI states in the system. [29]

**C. Current controlled helical light emission**

Having established efficient current control of spin/valley polarization, we now turn to a second key behavior in moiré MoTe$_2$ – spin/valley polarization dependent emission helicity. Although magnetism related phenomena have been observed in several moiré TMD systems, the spontaneous helical light emission is unique to tMoTe$_2$ because its bilayer form remains a direct bandgap semiconductor, while other TMD moiré systems become an indirect bandgap. Details of the emission mechanism have been discussed in depth in previous works. [6,10] In brief, spin/valley polarized carriers bind to electron-hole pair in the opposite valley to form singlet trions, which emit in the $\sigma^+$ ($\sigma^-$) channel for the electron-hole recombination in the +K (-K) valley. In the FM phase, due to spontaneous spin/valley polarization of the doped holes, light emission occurs only in a single valley, leading to circularly polarized emission at $\mu_0 H$ = 0. As shown below, current control of spin/valley polarization would also allow us to control trion emission helicity.

Figures 3(a)&(b) present the spatial maps of circular polarization resolved trion photoluminescence. The degree of circular polarization is defined as $\rho = \frac{PL(\sigma^-) - PL(\sigma^+)}{PL(\sigma^-) + PL(\sigma^+)}$. The data are taken with linearly polarized excitation at 632.8 nm, with $v$ = -0.77 and $I$ = $\pm 35$ nA - matching the

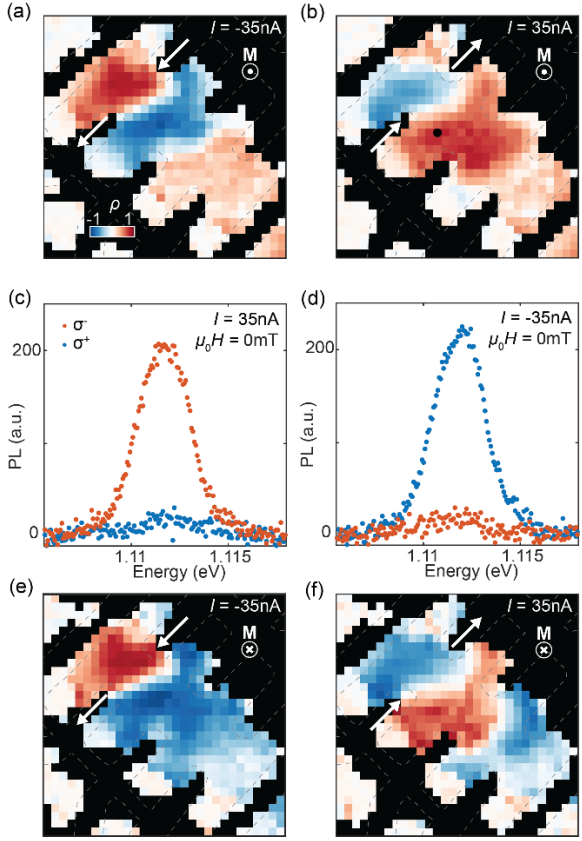

FIG 3. Current control of helical light emission. (a) Spatial map of PL degree of circular polarization $\rho$ with magnetization initialized to point up. Measurement taken with $I = -35$ nA current and in the anomalous Hall metal regime ($\nu = -0.77$) at $\mu_0 H = 0$ T. Domains with opposite signs of $\rho$ are visible on either side of the current channel. Finite, but smaller, $\rho$ is visible far from the current channel, on the lower-right side of the device. (b) Same as (a), but with $I = 35$ nA. The domain signs are reversed compared to (a). (c) Circular polarization-resolved PL spectra from the position marked by the black dot in (b). The PL shows clear helicity dependence of the light emission at $\mu_0 H = 0$T, with the signal predominantly in the $\sigma^-$ channel. (d) Same as (c), but with $I = -35$ nA, extracted from the same position in (a). The helicity dependence of the PL is switched, with signal now mainly in the $\sigma^+$ channel. (e) As in (a), but with magnetization initialized to point down. While the opposite sign of $\rho$ is visible far from the current channel on the lower-right of the device, the sign of the domains on either side of the current channel remains the same as in (a), with the same current flow direction but opposite magnetization. (f) As in (b), but with magnetization pointing down. Behavior of domains with switching of current and magnetization direction is consistent with RMCD behavior in Fig. 2.

experimental conditions of Figs. 2(c)&(d). In these spatial maps, domains with opposite signs of $\rho$ on either side of the current channel appear, which switch with reversal of $I$. This behavior demonstrates that the switching of the light helicity is caused by the reversal of spin/valley polarization by current, as demonstrated in our RMCD measurements in Fig. 2. Selecting a spot on one of these domains (black dot in Fig. 3(b)), we plot polarization resolved PL spectra (Figs. 3(c)&(d)) for $I = \pm 35$ nA at $\mu_0 H = 0$. The data clearly show that as we reverse the current flow direction, the trion luminescence switches helicity with near unity polarization. This establishes current control of light emission helicity. In addition, comparing the polarization resolved PL spatial maps between magnetization initialized up (Figs. 3(a)&(b)) and down (Figs. 3(e)&(f)), we observe the same domain behavior as for the RMCD maps in Fig. 2, as expected from the spin/valley Hall torque mechanism.

We next consider the emission helicity as a function of both applied magnetic field and current. As seen in Figure 4(a), $\rho$ vs $\mu_0 H$ swept down and up at $I = 0$ and $\nu = -0.77$ shows hysteretic behavior and finite $\rho$ at zero magnetic field. This is expected for a ferromagnetic system where near-unity $\rho$ arises from spin/valley polarization. The same measurement with injected current (Figs. 4(b)&(c)) shows that while hysteresis vs $\mu_0 H$ is still visible, the center of hysteresis moves away from $\mu_0 H = 0$, depending on the current direction. A full dependence of $\rho$ vs both $I$ and $\mu_0 H$ is shown in Figure 4(d). We observe that the $\rho$ can be continuously tuned as a function of both parameters. $\rho$ vs $I$ and $\mu_0 H$ for both magnetic field sweep directions is shown in FIG. S5. The tuning of the trion PL emission helicity as a function of current is highlighted in Figure 4(e) - it can be continuously varied between $\sigma^+$ and $\sigma^-$ at zero magnetic field.

Our experimental results can be compared to quantitative calculations using the spin diffusion equations in a slab geometry (Fig. 4(f)). [25,30] We define $\delta M_z$ as current-induced magnetization and $M_z^0 = g \mu_B n$ as sample magnetization. Here, $g$ is the g-factor of charge carriers, $\mu_B$ is the Bohr magneton, and $n$ is the carrier density. Using realistic sample parameters (see Methods), Fig. 4(g) plots the calculated spatial distribution of $\delta M_z / M_z^0$ for selected charge currents. The calculations show that $\delta M_z$ has opposite sign at opposite sides of the current channel, and becomes comparable to $M_z^0$ for a charge current of approximately 30 nA (Fig. 4(g)). This estimate is on the same order of magnitude as our experimental observation of current switching.

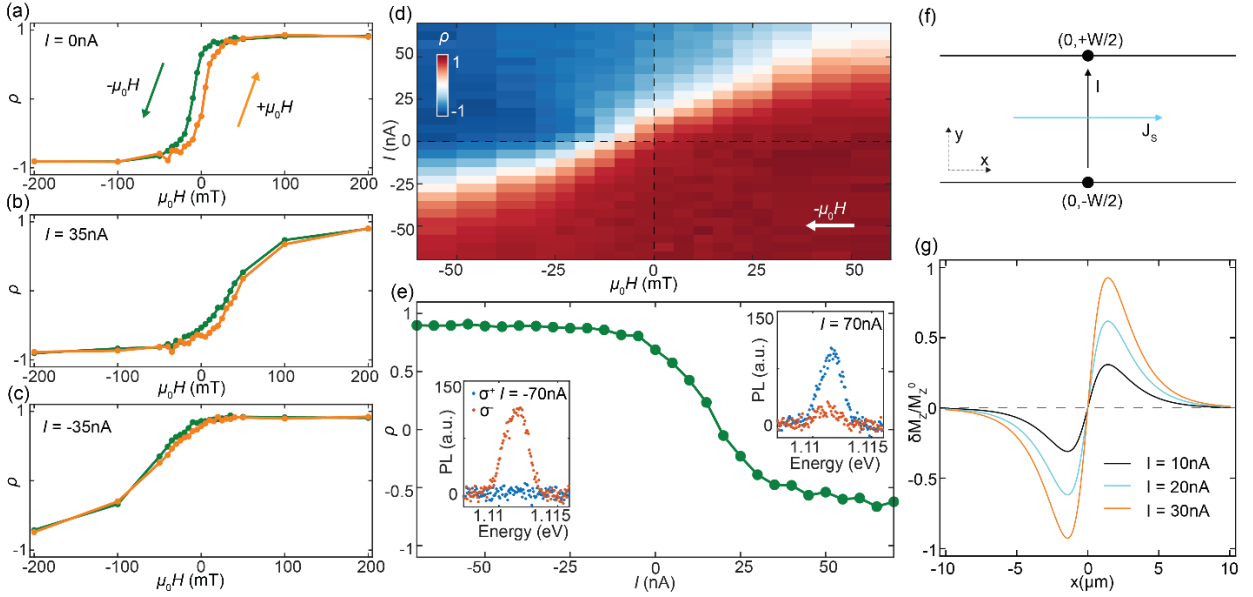

FIG. 4. Continuous tuning of emission helicity via current. (a) $\rho$ vs $\mu_0 H$ swept down and up, with $I = 0$nA. Hysteresis is visible, indicating that light emission helicity is controlled by the spin/valley polarization. (b)&(c) Same as (a), but with $I = 35$ nA (b) and $I = -35$ nA (c). While hysteresis of $\rho$ vs $\mu_0 H$ is still visible, applied current favors opposite spin/valley polarizations for opposite flow directions. (d) $\rho$ vs $I$ swept down, as a function of $\mu_0 H$, starting from positive $\mu_0 H$. Emission helicity can be continuously tuned between the $\sigma^+$ and $\sigma^-$ channels by both current and magnetic field. (e) $\rho$ as a continuous function of $I$, at $\mu_0 H = 0$T, initialized with $\mu_0 H > 0$ T. Insets: Helicity-resolved PL emission for large positive and negative current. Helicity can be continuously tuned between these two limits by changing $I$. All data are taken in the anomalous Hall metal regime, with $\nu = -0.77$, and with the beam spot on the opposite side of the current channel as for Figs. 3(c)&(d) (see FIG. S6). (f) Schematic of sample current flow configuration, with $I$ and $J_S$ injected charge and transverse spin currents, respectively. Current source and drain are at the black dots, with the current channel length W. (g) Induced magnetization $\delta M_Z$ versus $x$, derived from the spin diffusion equations and realistic sample parameters, at y=0. The critical charge current $I_C$ for magnetization flipping is about 30nA, the same order of magnitude as experimental observations (see Methods).

## III. CONCLUSIONS

Using RMCD and polarization resolved PL, we have established current control of spin/valley polarization and helical light emission in moiré MoTe$_2$. Although the device requires cryogenic temperatures, the strong coupling between electronic, spintronic, and photonic degrees of freedom demonstrated by these results could be appealing in a range of device applications. One possibility is to use these properties to couple magnetic information storage to optical communication within a single device. In addition, the topological phases which occur in tMoTe$_2$ could open new frontiers in topological opto-spintronics. As the integer and fractional Chern insulators recently observed in this system have a topological index which depends on the magnetization orientation, the observed spin/valley Hall torque implies that current injection can be used to manipulate these topological states. One interesting direction could be to use current injection to establish two magnetic domains with opposite orientations, and investigate the physics arising at the interface between two zero-field FCIs with opposite topological indices. We foresee that establishing current control of magnetic order in moiré MoTe$_2$ will add an important tuning knob to our toolbox as investigation of these zero-field FCIs and their anyonic excitations continues.

## IV. METHODS

### A. Device Fabrication

The moiré MoTe$_2$ samples equipped with electrical contacts used in this study were fabricated using the procedure discussed in more depth in Ref. 6. In brief, van der Waals flakes used in the heterostructure devices - graphite, hBN, and MoTe$_2$ - were mechanically exfoliated on oxygen-plasma cleaned Si/SiO$_2$ substrates and identified using an optical microscope. Atomic force microscopy was used to check hBN thickness. A graphite/hBN bottom gate with Pt contacts was prepared using conventional dry transfer, electron-beam lithography, e-beam evaporation, and contact-mode AFM cleaning techniques. In a glovebox with <0.1ppm H$_2$O and O$_2$ concentrations, monolayer MoTe$_2$ was exfoliated and an hBN-encapsulated moiré bilayer created with the cut-and-stack dry transfer technique, before putting down on the prepared backgate. After washoff and AFM cleaning of the device, a set of Pt contact gates and Au wire bonding pads were deposited using e-beam lithography and evaporation. Finally, the device was AFM cleaned again, before transferring on a top graphite gate.

### B. Transport Measurements

Transport measurements were taken in a Bluefors dilution refrigerator with a 9 T magnet and base electron temperature of about 80 mK. An AC current bias of 0.2–0.5 nA was generated using a 100 MΩ resistor in series with an AC voltage source (SR830), with the current monitored using a DL1211 current amplifier. Four-terminal $R_{xx}$ and $R_{xy}$ signals were amplified using the differential mode of an SR560 voltage preamplifier with an input resistance (about 100 MΩ) much larger than the contact resistance of the device. The amplified voltage signals were demodulated and measured using SR830/SR860 lock-ins.

### C. Optical Measurements

Optical measurements were taken in a closed-loop magneto-optical cryostat (attoDRY 2100) with an attocube $xyz$ piezo stage and $xy$ scanners, 9 T z-axis superconducting magnet, and with a base temperature of 1.6 K. Current was applied to the sample using a 100 MΩ resistor in series with a DC voltage source (Keithley 2450) connected to the source pin, with the drain pin grounded. Contact gates were set to -3 V to ensure transparent electrical contact to the moiré MoTe$_2$, as discussed in Ref. 6. Polarization-resolved photoluminescence measurements were taken with linearly polarized 632.8 nm HeNe laser excitation focused on the sample by a high-NA nonmagnetic cryogenic objective to a ~1 μm beam spot. Sample emission was collected by the same objective and passed through a quarter wave plate and linear polarizer to select out right and left circularly polarized channels. Signal was then passed through 75 μm pinhole and dispersed with a diffraction grating (Princeton Instruments, 600 grooves/mm at 1 μm blaze) and detected by a LN cooled infrared CCD (Princeton Instruments PyLoN-IR 1.7).

RMCD measurements were taken with excitation near the trion resonance by filtering a broadband supercontinuum source (NKT SuperK Fianium FIR-20) by dual-passing through a monochromator to achieve a narrow excitation bandwidth. The out-of-plane magnetization of the sample induces a MCD signal ΔR, the difference between the reflected right- and left-circularly polarized light. To obtain the normalized RMCD ΔR/R, the laser intensity was chopped at $p$ = 850 Hz and the phase was modulated by λ/4 via a photoelastic modulator at $f$ = 50 kHz. An InGaAs avalanche photodiode detector was used to collect the reflected signal, and the output was read by two lock-in amplifiers (SR830). The ratio between the $p$-component signal $I_p$ and $f$-component signal $I_f$ gives the RMCD signal: $\varDelta R/R = I_f /(J_1(\pi/2) \times I_p)$ where $J_1$ is the first-order Bessel function.

### D. Determination of doping density and electric field

A parallel plate capacitor model was used to determine the carrier density $n$ and electric field $D$ from the applied top and bottom gate voltages, $V_{tg}$ and $V_{bg}$. Gate capacitances $C_{tg}$ and $C_{bg}$ are calculated using the hBN thickness determined by atomic force microscopy, taking the hBN dielectric constant to be 3.0. Thus, $n$ and $D$ can be computed as $n = (V_{tg}C_{tg} + V_{bg}C_{bg})/e − n_{\text{offset}}$ and $D/\varepsilon_0 = (V_{tg}C_{tg} − V_{bg}C_{bg})/2\varepsilon_0 − D_{\text{offset}}$, with $e$ the electron charge and $\varepsilon_0$ the vacuum permittivity. Carrier density offset $n_{\text{offset}}$ is derived from fitting to the insulating states in the PL spectra and transport measurements. $D_{\text{offset}}$ is determined from the symmetry axis of the RMCD phase diagram.

### E. Induced magnetization and spin Hall current

The magnetization induced by the spin Hall currents can be derived from the spin diffusion equations. [25,30] The configuration considered is shown in Fig. 4(f). A charge current $I$ is injected in the y direction into a slab of width $W$, generating a spin current $J_S$ in the x direction to flip the magnetization. Solving the diffusion equation in the slab, the induced magnetization [25,30] is

$$\delta M_z(x,y) = \frac{2g\mu_B I \tan\theta_{SH}}{e l_S^2/\tau_S} \int \frac{dk}{2\pi} e^{ikx} \frac{\cosh(\omega_k y)}{\omega_k \coth(kW/2)\sinh(\omega_k W/2) + 4k\tan^2(\theta_{SH})\cosh(\omega_k W/2)},$$

where $g$ is the g factor of charge carriers, $\mu_B$ is the Bohr magneton, $\theta_{SH}$ is the spin Hall angle, $e$ is the electron charge. $l_S$ is the spin diffusion length, and $\tau_S$ is the spin relaxation time. $\omega_k = \sqrt{k^2 + l_S^{-2}}$.

The calculation parameters are estimated from experimental values. $W = 3\ \mu m$ is the distance between electrical contacts. To estimate the value of $l_S$, we note that experimentally the size of the flipped domain is on the order of $\mu m$, indicating $l_S$ should be on the same order. We choose $l_s = 1\ \mu m$. The spin Hall angle is defined by $\tan\theta_{SH} = e\sigma_{SH}/\hbar\sigma_{xx}$, where $\sigma_{SH}$ is the spin Hall conductivity, $\sigma_{xx}$ is the longitudinal conductivity and $\hbar$ is the reduced Planck constant. $\sigma_{SH} = (\sigma_{H,K} - \sigma_{H,K'})\hbar/2e$ (Ref. [31]) due to spin-valley coupling in TMDs, with $\sigma_{H,K}$ and $\sigma_{H,K'}$ the Hall conductivities of valleys K and K'. If we assume the system is fully polarized to K, $\sigma_{SH}$ is simply $\sigma_{H,K}\hbar/2e = \hbar\sigma_{AH}/2e$ and $\sigma_{AH}$ is the anomalous Hall conductivity. Thus, $\tan\theta_{SH} = \sigma_{AH}/2\sigma_{xx}$. The experimentally measured longitudinal resistance $\rho_{xx}$, and anomalous Hall resistance $\rho_{xy}$ are both on the same order of a few kΩ. We thus take $\tan\theta_{SH} = \sigma_{AH}/2\sigma_{xx} = \rho_{xy}/2\rho_{xx} = 1/2$. Finally, while $\tau_s$ cannot be determined directly from our measurement, $\tau_s$ in TMD heterobilayers is estimated to be on the order of $\mu s$ in Refs. [25,32]. We take $\tau_s$ to be on the same order here, $\tau_s = 5\mu s$.

Fig. 4(g) shows the induced magnetization $\delta M_z$ vs position $x$ at the center of the slab ($y = 0$) for different injection currents $I$. The magnetization flipping occurs when the peak of $\delta M_z$ is on the order of $M_z^0$, where $M_z^0$ is the magnetization of the ground state. $M_z^0 = g\mu_B n$, and carrier density $n = -3.8 \times 10^{12}\ cm^{-2}$ for the filling $\nu = -0.78$ in our experimental sample. With our choice of $\tau_s$, the critical current for magnetization flipping is on the order of 30 nA, consistent with the experimental values.

## ACKNOWLEDGMENTS

This project is mainly supported by the U.S. Department of Energy (DOE), Office of Science, Basic Energy Sciences (BES), under the award DE-SC0018171. Electrical control of ferromagnetism is partially supported by Vannevar Bush Faculty Fellowship (Award number N000142512047). Theory is supported by DE-SC0012509. E.A. acknowledges support by the National Science Foundation Graduate Research Fellowship Program under grant no. DGE-2140004. The authors also acknowledge the use of the facilities and instrumentation supported by NSF MRSEC DMR-1719797. K.W. and T.T. acknowledge support from the JSPS KAKENHI (Grant Numbers 21H05233 and 23H02052) , the CREST (JPMJCR24A5), JST and World Premier International Research Center Initiative (WPI), MEXT, Japan. XX acknowledges support from the State of Washington funded Clean Energy Institute and from the Boeing Distinguished Professorship in Physics.

XX conceived and supervised the experiment. HP fabricated and performed transport measurements, assisted by JC. EA performed the magneto-optical measurements. EA, XX, DX, TC, and LF analyzed and interpreted the results. KY and DX performed calculations. TT and KW synthesized the hBN crystals. EA, XX, and DX wrote the paper with input from all authors. All authors discussed the results.

The authors declare no competing interests.

## DATA AVAILABILITY

All supporting data for this paper and other findings of this study are available from the corresponding author upon reasonable request.

# Supplemental Material for

## Magnetoelectric Control of Helical Light Emission in a Moiré Chern Magnet

Eric Anderson[1*], Heonjoon Park[1], Kaijie Yang[2], Jiaqi Cai[1], Takashi Taniguchi[3], Kenji Watanabe[4], Liang Fu[5], Ting Cao[2], Di Xiao[2,1†], and Xiaodong Xu[1,2‡]

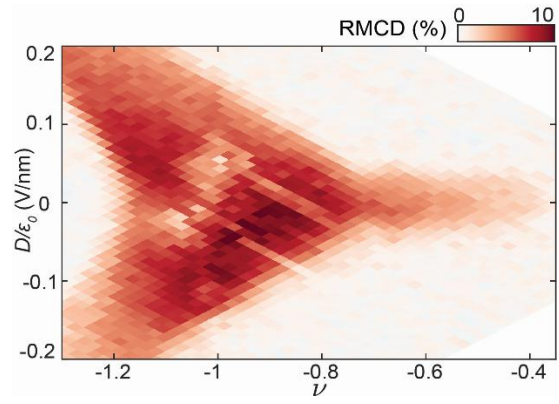

FIG. S1. RMCD at zero magnetic field. RMCD measurement vs $\nu$ and $D/\varepsilon_0$ in D(3.7°), taken at $\mu_0H = 0$ T and 1.6K

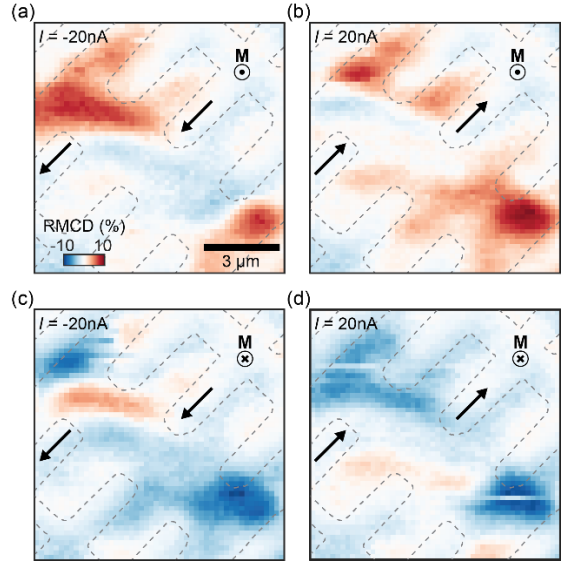

FIG. S2. Alternate current channel RMCD maps. (a) RMCD spatial map in the anomalous Hall metal regime ($v = -0.77$) with magnetization initialized to point up. $I = -20$ nA current flows between the contacts on the bottom left and center top pins. Grey dashed lines indicate Pt contact area. Similar behavior to FIG. 2(c) is observed, but with a different current channel due to the different pins. (b) As in (a), but with a $I = 20$ nA current flowing between the contacts. (c)&(d), As in (a)&(b), but with magnetization initialized to point down.

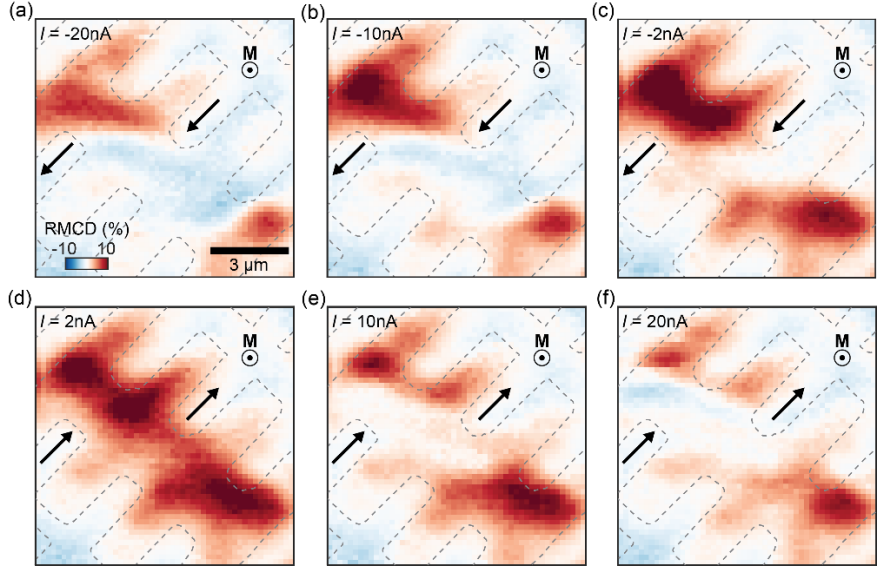

FIG. S3. RMCD maps at selected currents. All maps were taken with magnetization initialized to point up. (a) As in FIG. S2(a): RMCD spatial map in the anomalous Hall metal regime ($v = -0.77$) with a $I = -20$ nA current. (b) As in (a), but with a $I = -10$ nA current. Sign reversed domain is visible below the current channel. (c) As in (a)&(b), but with $I = -2$nA current. Even this small value of current affects the sample spin/valley polarization, with the RMCD map clearly different from that with no current flowing, as shown in FIG. 2(b). (d)-(f), as in (a)-(c), but with currents flowing in the opposite direction of 2 nA, 10 nA, and 20 nA, respectively.

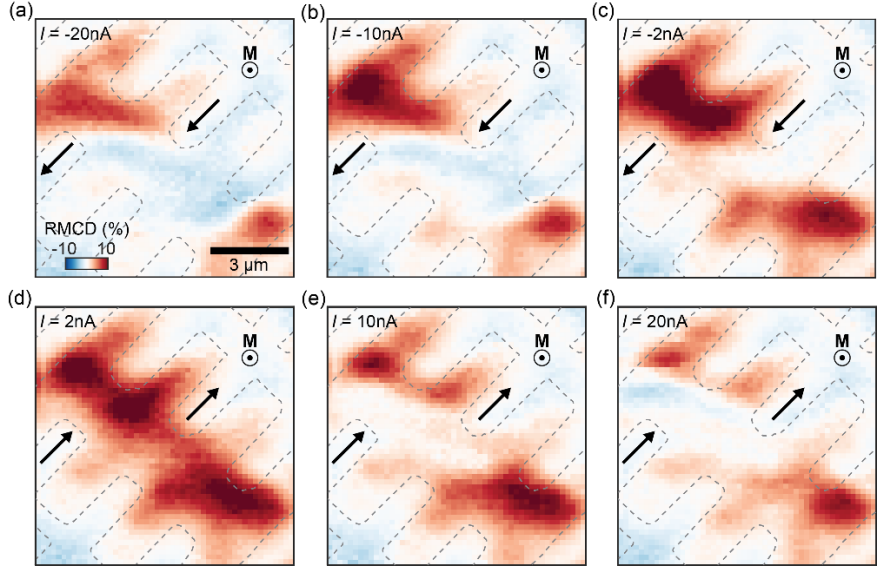

FIG. S3. RMCD maps at selected currents. All maps were taken with magnetization initialized to point up. (a) As in FIG. S2(a): RMCD spatial map in the anomalous Hall metal regime ($v = -0.77$) with a $I = -20$ nA current. (b) As in (a), but with a $I = -10$ nA current. Sign reversed domain is visible below the current channel. (c) As in (a)&(b), but with $I = -2$nA current. Even this small value of current affects the sample spin/valley polarization, with the RMCD map clearly different from that with no current flowing, as shown in FIG. 2(b). (d)-(f), as in (a)-(c), but with currents flowing in the opposite direction of 2 nA, 10 nA, and 20 nA, respectively.

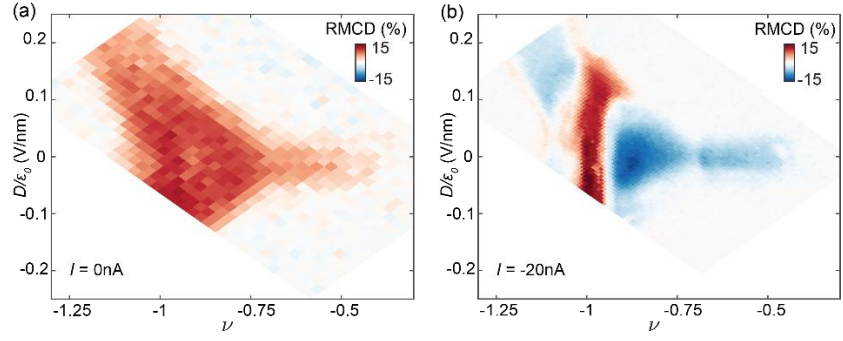

FIG. S4. Current switching of spin/valley polarization across the flat moiré Chern band. (a) RMCD vs $v$ and $D/\varepsilon_0$ in D(3.9°), with no current flowing, consistent with previous measurements. (b) As in (a), but with $I = -20$ nA flowing between the contacts. RMCD is measured on the same spot as FIG. 3. Clear sign change is observed throughout the anomalous Hall metal phase space of the partially filled Chern band, as well as in the anomalous Hall metal tails above $v = -1$ and at finite $D/\varepsilon_0$.

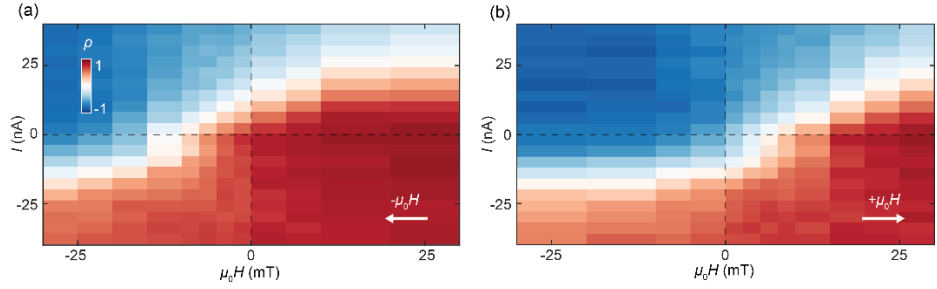

FIG. S5. Emission helicity vs magnetic field sweep direction. (a) $\rho$ vs $I$ swept down, as a function of decreasing magnetic field $\mu_0 H$, starting from positive $\mu_0 H$. (b) As in (a), but with increasing $\mu_0 H$ starting from starting from negative $\mu_0 H$. Opposite signs $\rho$ of at $I = 0$ nA and $\mu_0 H = 0$ mT show hysteresis as a function of magnetic field sweep direction. Data taken on the same spot as FIG. 4.

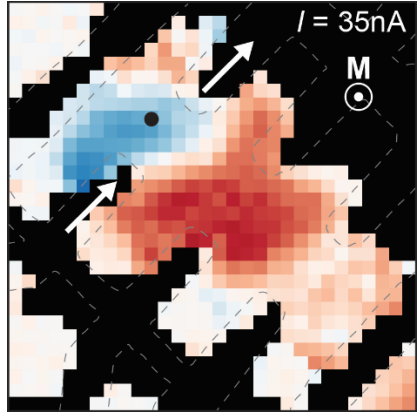

FIG. S6. Measurement location of FIG. 4 in the main text. Spatial map of PL degree of circular polarization $\rho$ with magnetization pointing up. Measurement taken with $I = 35$ nA current and in the anomalous Hall metal regime $\nu = -0.77$, at $\mu_0 H = 0$ T, after magnetization initialization at $\mu_0 H > 0$ T. Black dot denotes location of data in FIG. 4.